\documentclass[10pt, conference, compsocconf]{IEEEtran}

\usepackage[utf8]{inputenc}
\usepackage[english]{babel}
\usepackage{gensymb}

\usepackage{multirow, multicol, booktabs, tabulary, tabu, longtable, array, varwidth}
\usepackage{placeins, lipsum}
\usepackage[flushleft]{threeparttable}

\usepackage[labelfont=bf]{caption}
\usepackage{cite}
\usepackage[hyphens]{url}
\usepackage{hyperref}
\usepackage[dvipsnames, table]{xcolor}
\hypersetup{
    linkcolor  = blue!85!black,
    citecolor  = blue!85!black,
    urlcolor   = blue!85!black,
    colorlinks = true,
    breaklinks = true
}

\usepackage{enumerate}

\usepackage[inline]{enumitem}

\usepackage{amsmath, amsfonts, amssymb, amsthm, nicefrac}
\usepackage{mathptmx}  
\usepackage[mathcal]{eucal}

\theoremstyle{definition}
\newtheorem{definition}{Definition}

\theoremstyle{comment}

\newtheorem*{proposition}{Proposition}

\usepackage[]{cleveref} 
\crefname{appsec}{Appendix}{Appendices}
\crefformat{section}{\S#2#1#3}
\crefformat{subsection}{\S#2#1#3}
\crefformat{subsubsection}{\S#2#1#3}
\crefformat{equation}{(#2#1#3)}
\Crefformat{equation}{Equation~(#2#1#3)}
\crefrangeformat{equation}{(#3#1#4--#5#2#6)}
\Crefrangeformat{equation}{(#3#1#4--#5#2#6)}
\crefmultiformat{equation}{(#2#1#3)}{ and~(#2#1#3)}{, (#2#1#3)}{ and~(#2#1#3)}
\crefformat{figure}{Fig.~#2#1#3}
\Crefformat{figure}{Fig.~#2#1#3}
\crefmultiformat{figure}{Figs.~#2#1#3}{ and~#2#1#3}{, #2#1#3}{ and~#2#1#3}
\crefformat{algorithm}{Algorithm~#2#1#3}
\Crefformat{algorithm}{Algorithm~#2#1#3}
\crefmultiformat{algorithm}{Algorithm~#2#1#3}{ and~#2#1#3}{, #2#1#3}{ and~#2#1#3}
\crefformat{table}{Table~#2#1#3}
\Crefformat{table}{Table~#2#1#3}
\crefrangeformat{table}{Tables~#3#1#4--#5#2#6}
\Crefrangeformat{table}{Tables~#3#1#4--#5#2#6}
\crefmultiformat{table}{Tables~#2#1#3}{ and~#2#1#3}{, #2#1#3}{ and~#2#1#3}

\usepackage{textcomp}
\usepackage[shortcuts,acronym]{glossaries}
\usepackage{soul, footnote, xargs}
\usepackage[colorinlistoftodos,prependcaption,textsize=tiny]{todonotes}
\usepackage{pifont}
\usepackage{balance}

\usepackage[scaled=0.82]{beramono}

\usepackage{graphicx}
\usepackage[tight,footnotesize]{subfigure}
\usepackage{tikz, epstopdf, stfloats, bbding, capt-of}
\usepackage{algorithmic}
\usepackage[ruled, linesnumbered]{algorithm2e}

\SetCommentSty{mycommfont}
\graphicspath{{figs/}}
\DeclareMathOperator*{\argmax}{arg\,max}
\renewcommand{\footnotesize}{\scriptsize}
\newcommand{\quotes}[1]{``#1''}

\usepackage{wasysym}

\usepackage{microtype}
\usepackage[T1]{fontenc}
\def\baselinestretch{0.93}

\makeatletter
\newcommand\subparagraph{%
    \@startsection{subparagraph}{5}
    {\parindent}
    {3.25ex \@plus 1ex \@minus .2ex}
    {-1em}
    {\normalfont\normalsize\bfseries}}
\makeatother
\usepackage[compact]{titlesec}
\let\subparagraph\relax 
\titlespacing\section{0pt}{6pt plus 4pt minus 2pt}{2pt plus 2pt minus 2pt}
\titlespacing{\subsection}{0pt}{4pt plus 2pt minus 1pt}{2pt plus 1pt minus 1pt}
\titlespacing{\subsubsection}{0pt}{4pt plus 2pt minus 1pt}{2pt plus 1pt minus 1pt}
\usepackage{etoolbox}
\makeatletter
\patchcmd{\ttlh@hang}{\parindent\z@}{\parindent\z@\leavevmode}{}{}
\patchcmd{\ttlh@hang}{\noindent}{}{}{}
\makeatother

\begin{document}
\title{On Preempting Advanced Persistent Threats Using Probabilistic Graphical Models} 

\author{
    \IEEEauthorblockN{Phuong Cao}
    \IEEEauthorblockA{University of Illinois at Urbana-Champaign}
}

\maketitle
\begin{abstract}
This paper presents PULSAR, a framework for 
preempting Advanced Persistent Threats (APTs). PULSAR employs a probabilistic
graphical model (specifically a Factor Graph) to infer the time evolution of an attack based on observed security events at runtime. The framework 
\begin{enumerate*}[label=(\roman*)]
    \item learns the statistical significance of patterns of events from past attacks;
    \item composes these patterns into FGs to capture the progression of the attack; and
    \item decides on preemptive actions.
\end{enumerate*}
The accuracy of our approach and its performance are evaluated in three experiments at SystemX:
\begin{enumerate*}[label=(\roman*)]
    \item a study with a dataset containing 120 successful APTs over the past 10 years (PULSAR accurately 
    identifies 91.7\%);
    \item replaying of a set of ten unseen APTs (PULSAR stops 8 out of 10 replayed
    attacks before system integrity violation, and all ten before data exfiltration); and
    \item a production deployment of the framework (during a month-long deployment, PULSAR took an 
    average of one second to make a decision).
\end{enumerate*}

\end{abstract}

\begin{IEEEkeywords}
    Factor Graphs, Attack Preemption
\end{IEEEkeywords}

\section{Introduction}
\emph{Advanced Persistent Threats} (APTs) are among the most sophisticated attacks targeting
networked systems~\cite{NIST-APT}. Instead of exploiting a single vulnerability, an APT consists of
several stages: an attacker
\begin{enumerate*}[label=(\roman*)]
    \item gains unauthorized access to a network,
    \item uses multiple attack vectors to pursue objectives repeatedly~\cite{ibmxforce17, attck,Lauinger2017ThouSN}, and
    \item remains undetected for extended periods of time by staying under the radar of monitors\cite{aptnotes}.
\end{enumerate*}
Network and host security monitors~\cite{paxson1999bro} generate security events in APTs that often
overlap with legitimate user activities. These result in high false positive rates (FPR)~\cite{axelsson1999base,basso2010analysis} by
threat detection software (TDS)~\cite{roesch1999snort} and security information and event
management tools (SIEM)~\cite{bray2008ossec}.

Despite the deployment of TDS/SIEM in large networks, successful attacks occur because security
events by themselves (particularly if considered in isolation) are not sufficient indicators of
malicious behavior or its progression. Often, system integrity has already been compromised (e.g.,
a rootkit is already installed) or data have been exfiltrated~\cite{equifax,muthukumaran2015flowwatcher} by the time a critical
event occurs. Today, such critical events can only be handled by a nimble team of experienced
analysts~\cite{d2008real} using forensics and provenance tools~\cite{bates2015trustworthy,nodoze}
among others. While these tools are useful, analysts still rely on their domain knowledge and
learned (based on past security incidents) data characteristics (e.g., repetitiveness, severity,
and common patterns of events) together, with the ongoing observation of events to investigate or
respond to attacks.

The goal of this paper is to preemptively stop APTs with
minimal false positives and low performance overhead. We use machine learning to fuse domain
knowledge, experience of past attacks, and real-time observations from security monitors to detect
and preempt an ongoing attack. We demonstrate this approach (PULSAR) in a production
environment with real data from a large-scale, high-performance computing infrastructure that has 
thousands of nodes and both academic and industrial users (anonymized under the name SystemX). The specific approach is
based on probabilistic graphical models (PGM), particularly Factor Graphs (FG). The probabilistic
nature of the model captures the uncertainty in
\begin{enumerate*}[label=(\alph*)]
    \item handling incomplete data and, 
    \item handling the significance of an event under different circumstances (as shown 
    in \cref{sec:modeling}).
\end{enumerate*}
In addition, FGs allow us 
\begin{enumerate*}[label=(\roman*)]
    \item to capture temporal relationships among observed events,
    \item to build compositional models from the captured relationships to allow for machine 
    interpretability, and
    \item to use the above information to form scalable real-time inference strategies so as to 
    preempt attacks.
\end{enumerate*}
We assert that FGs are a suitable formulation for modeling security attacks, because FGs model
conditional dependencies and provide mechanisms to incorporate domain knowledge, while operating on
modest data sizes (e.g., N = 120 attacks in this paper) with class
imbalance. FGs are suitable for security analysis because
successful attacks are rare (e.g., they represent only 0.01\% of events in
SystemX's daily operation). An important benefit of FGs in contrast to provenance models~\cite{bates2015trustworthy,nodoze} is that relationships
between sequences of security events and attack stages can be modeled statistically from historical
data, and no assumptions are made about the causal ordering among events.

\textbf{Contributions.} The key contributions of this paper are:
\begin{enumerate*}[label=(\roman*)]
    \item An FG to model the progress of APTs targeting networked systems with thousands of nodes.
    \item A method to learn FG from annotated real-word attacks in which 99.7\% of the data is 
    annotated automatically. 
    \item A testbed for testing ten attacks unseen at SystemX in the presence of 
    legitimate traffic.
    \item A comprehensive runtime and accuracy evaluation. 
\end{enumerate*}

\textbf{Evaluation.}
We demonstrate PULSAR's accuracy and performance via three experiments. \emph{First}, we evaluated PULSAR on 120 real APTs
observed in the past 10 years at SystemX. The dataset contains a variety of APTs that use different types of attack vectors, exploits, and payloads. When we trained PULSAR on one half of the historical attacks (60 out of 120 APTs) and tested it on the other half, PULSAR accurately
identified 91.7\% APTs prior to data loss. 
\emph{Second}, to test its generalizability, we trained PULSAR on the entire historical dataset of 120 known APTs and tested it on ten APTs, unseen at SystemX. The unseen APTs, described in \cref{tab:scenarios}, are based on top ten attack techniques described in the IBM Threat
Intelligence Index~\cite{ibmxforce17}. In order to have realistic attack scenarios, the unseen APTs were injected into live traffic where background events intermingled with attack events. PULSAR successfully stopped eight out of the ten unseen APTs before system integrity violation and all the APTs before data loss. \emph{Third}, PULSAR was integrated into SystemX's security infrastructure for a month-long deployment. During this period, there were an average of 94,238 events per day. PULSAR filtered this stream of events to an average of 1,885 significant events that it performed inference on at an average rate of one second per event.

\textbf{Putting PULSAR in Perspective.}
While traditional SIEMs can output events of related attack vectors, such events often overwhelm
security analysts~\cite{cao2014abusing,nodoze,cotroneo2016automated,paudice2014experiment}. Anomaly-based techniques
(e.g.,~\cite{xu2016sharper,yen2013beehive,shon2005machine}) have potential to capture novel APTs,
but they require extensive observations of normal usage profiles to detect anomalous activities, and also need substantial tuning to minimize FPR. Recent provenance-based techniques~\cite{nodoze,milajerdiholmes} aim to build
dependency graphs of APTs, however, these techniques require a complete causal observation of an APT
and may be more suitable for offline analysis~\cite{bates2015trustworthy}. Our approach is
unique in that i) it makes no assumptions about the causal ordering among events, and ii) it is trained and validated on both longitudinal and live production traffic.
\begin{table}[!t]
    \centering
    \caption{Listing of attack stages $\sigma_i \in \mathcal{S}$.}
    \rowcolors{2}{gray!15}{white}
    \resizebox{\columnwidth}{!}{%
        \begin{tabular}{lp{2.2cm}p{4cm}}
            \toprule
            \textbf{Stage} & \textbf{Name} & \textbf{Description/Example} \\
            \midrule
            $\sigma_0$& Benign & Legitimate uses of the target system \\
            $\sigma_1$& Discovery & Scan for open ports or applications\\
            $\sigma_2$& Initial Access & Remote login into the target system\\
            $\sigma_3$& Gathering & Reading kernel version\\
            $\sigma_4$& Command \& Control & Receiving attacker commands \\
            $\sigma_5$& Preparation & Obtaining and compiling exploits\\
            $\sigma_6$& Persistence & Installation of backdoor\\
            $\sigma_7$& Lateral Movement & Accessing internal hosts\\
            $\sigma_8$& Defense Evasion & Purging attack traces \\
            $\sigma_9$& Collection & Recording secret keys or passwords\\
            $\sigma_{10}$& Exfiltration & Using covert channels to extract secrets\\
            \bottomrule
        \end{tabular}%
    }
    \label{tab:stages}
\end{table}

\section{Overview}
\label{sec:overview}
\subsection{Preliminaries}
\begin{definition} \label{def:event}
A \emph{security event} (also referred to as an event) $e_t \in \mathcal{E}$ is a variable that represents
an observation of a potentially malicious activity at a time $t$ using one or more security-related
log messages from a TDS/SIEM. The set $\mathcal{E}= \{ \epsilon_1, \dots \epsilon_m \}$ contains 105
possible values of events indicating APT activities found in both SystemX and other similar systems~\cite{badger2015scalable}.
\end{definition}

\begin{definition}
An \emph{attack stage} $s_t \in \mathcal{S}$ represents the progression of an APT
at time $t$. The set $\mathcal{S}=\{\sigma_0, \dots \sigma_{n-1} \}$ contains $n$ possible values of
attack stages. This paper adopts the MITRE classification~\cite{attck} that defines $n=11$ stages
(see \cref{tab:stages}) commonly found in APTs in the wild. Publicly disclosed APTs~\cite{aptnotes}
follow these attack stages and have been used in prior work on APTs~\cite{milajerdiholmes}.
\end{definition}

\subsection{Severity, Repetitiveness \& Commonality}\label{sec:sev_rep_com}
The key challenge for a security analyst is to identify an event or a set of events that are
good indicators of an APT before it has caused any damage. To characterize known APTs, we analyze a
longitudinal dataset (2008--present) of 120 security incidents at SystemX. Our dataset includes 
\begin{enumerate*}[label=(\roman*)]
    \item human-written incident reports that indicate the users and the machines involved in the 
    incident,
    \item raw logs of both legitimate user activities and attack activities, i.e., network flows
    (generated by a cluster of \texttt{Bro} network security monitors (NSM)~\cite{paxson1999bro}),
    system logs (generated by \texttt{rsyslog,osquery,} and \texttt{ossec}~\cite{bray2008ossec}), and
    \item audit logs of system calls (generated by \texttt{auditd}).
\end{enumerate*}
These raw logs contain detailed attack activities. For each log message, we wrote scripts to remove specific (e.g., personal information~\cite{talbi2018towards,chen2017evaluating} or IP addresses) and non-deterministic information (e.g., time) and keep only the event in
$\mathcal{E}$ in the form of a symbolic name. For example, \texttt{download\_sensitive} denotes a
download of a file with a sensitive extension, e.g., \texttt{.c}, from an arbitrary node at any time. We
assert that our dataset captures variants of known APTs and generalizes to unseen APTs (see~\cref{sec:eval}).

In our dataset of 235K events, a majority of events (99.7\%) have been automatically annotated with
corresponding attack stages. These events are clearly benign (e.g., \texttt{login}) or clearly
malicious (e.g., installation of a binary file in an existing malware database). Only a small fraction
(0.3\%) of events (i.e., ones that appear in both attack and legitimate activities) cannot be
annotated automatically. We consulted with several security experts to annotate the remaining
events. While we assume that the annotations by security experts are correct, i.e., attack events
are labeled as malicious, we can reuse a body of work in ML that addresses annotation
accuracy~\cite{karger2011iterative,ratner2017snorkel}. Below we discuss the main characteristics of our dataset on APTs.

\begin{figure}[!t]
    \centering
    \includegraphics[width=0.45\textwidth]{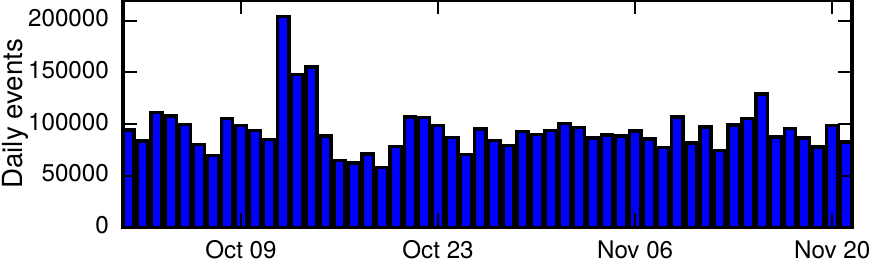}
    \caption{SystemX's monitors observe an average of 94,238 events per day (standard deviation = 23,547) in Oct--Nov 2018.}
    \label{fig:alerts-per-month}
\end{figure}

\textbf{Severity.}
\emph{Severe events} can be used to detect successful APTs; however, they cannot be used to preempt
attacks because their occurrences indicate that the system integrity has already been
compromised and that data have already been exfiltrated~\cite{bonaci2015experimental}. In fact, the entire
dataset has 19 such unique critical events which occur 98 times in the 120 APTs. In all cases those
critical events were detected when it was too late to preempt the system integrity loss. On the other
hand, if any of events was considered as an indicator of a  complete APT then analysts would have to analyze all of low- and medium-severity events (e.g., $94K$ daily events
observed at SystemX in \cref{fig:alerts-per-month}). 

\begin{figure}[!t]
    \centering
    \includegraphics[width=\columnwidth]{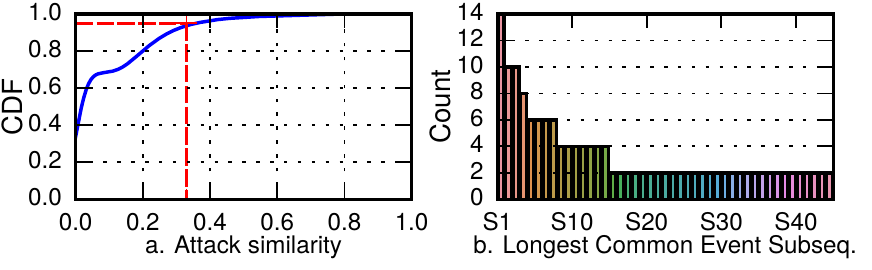}
	\caption{(a) The fractions of similar events between pairs of APTs in our dataset. (b) The count of LCS in our dataset.}
    \label{fig:attack_similarities}
\end{figure}

\textbf{Commonality.}
Another way of identifying APTs is to look at characteristics shared by already known and new APTs, e.g., the longest common subsequence (LCS)~\cite{nistlcs} of events that lead
up to malicious activity. For example, in
\cref{fig:attack_similarities}a, we observe that 95\% of the attacks in our dataset share at most
33.3\% of their events. These events correspond to common attack vectors for establishing a foothold
in the target network before executing exploits to exfiltrate secrets.
\cref{fig:attack_similarities}b shows the histogram of 45 LCS (S1--S45) identified in our dataset.
The histogram indicates that LCSs have common patterns across multiple attacks - which can be
learnt.

\textbf{Repetitiveness.}
APTs observed in our dataset (and in the wild~\cite{aptnotes}) often start with a set of
\emph{repetitive} but \emph{inconclusive} events to identify vulnerable computing resources (e.g.,
scans for vulnerable Apache Struts portals~\cite{equifax}). Such repeated events are themselves not
indicators of malicious activity, but can be used to signal potentially malicious events that need further monitoring. SystemX observes an average of 80k (out of 94K in \cref{fig:alerts-per-month}) repeated port and
vulnerability scans on a daily basis.

\begin{figure}[!t]
    \centering
    \includegraphics[width=\columnwidth]{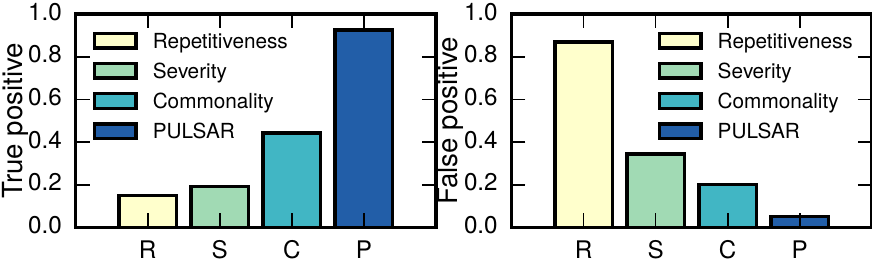}
	\caption{Combination of multiple events (using PULSAR) provides more accurate detection.}
    \label{fig:apt_csr}
\end{figure}

To summarize, use of individual events categorized by only one of the three data characteristics
(i.e., severity, commonality, or repetitiveness) can detect the presence of an APT, but with two
caveats:
\begin{enumerate*}[label=(\roman*)]
    \item the APT is already successful (i.e., the system integrity is already lost), or
    \item the FPR is high.
\end{enumerate*}
We find, in the historical data at SystemX, that using events
categorized by a single data characteristic can only detect 44\% (53 out of 120) past APTs. Moreover, using individual events resulted in a high FPR of 87\% when using only repetitive events as
indicators of APTs as shown in~\cref{fig:apt_csr}.

\begin{proposition}
We assert that combining events from diverse monitors with different characteristics and jointly
analyzing (fusing) their statistical significance (likelihood of belonging to an attack) can lead to a more
accurate APT detector that can effectively preempt attacks.
\end{proposition}

We show how PULSAR works with a sophisticated attack.

\subsection{Motivating Example: The Attack} \label{sec:attackers_perspective}
The Equifax attack 
\begin{enumerate*}[label=(\roman*)]
    \item used a remote execution exploit (RCE) (publicly disclosed as
    CVE-2017-5638~\cite{CVE-2017-5638}) to gain access;
    \item stayed undiscovered in Equifax's system for an extended period
    of time~\cite{equifax};
    \item extracted 143M Social Security Numbers.
\end{enumerate*}
The goal of the attack was to get a shell (terminal) to control the Struts server without
compromising a user's account. While the CVE of the remote exploit is publicly known, we do not have
the specifics of steps (ii) and (iii). In our attack, we used 
\begin{enumerate*}[label=(\roman*)]
    \item a remote code execution exploit targeting Apache Struts (CVE-2017-5638, identical to the
    vulnerability in the Equifax breach),
    \item a privilege escalation (PE) exploit~\cite{CVE-2016-5195},
    \item a rootkit that uses port-knocking for stealthy data extraction using a DNS tunnel.
\end{enumerate*}
To simulate the attack, we collaborated with SystemX's redteam to setup a vulnerable Struts
server in a virtual machine located in a production cluster at SystemX and launch the attack
(i.e., via the RCE) from outside of SystemX. We collaborated with the SystemX
analysts to analyze the event streams so as
to understand their thought process.

\begin{figure*}[!t]
    \centering
    \includegraphics[width=\textwidth]{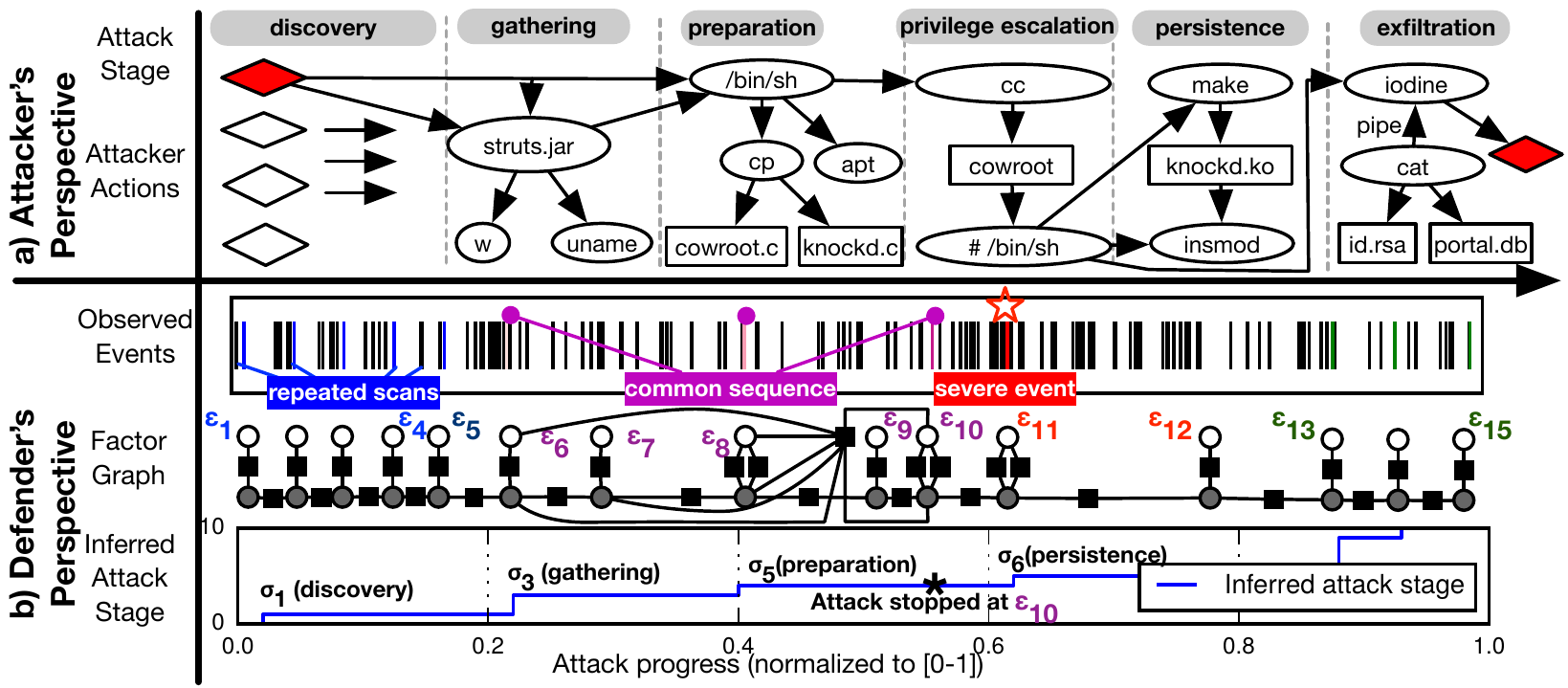}
    \caption{(a) Attacker's view of the exercised APT. (b) Defender's view of detected attack stages and recommended actions.}
    \label{fig:approach_overview_data}
\end{figure*}

While most APTs can take days to run, it would have been impractical for our attack to disrupt SystemX's
production systems for an extended period of time. Thus, we ran the attack for 10 minutes in a
24-hour period while production workloads were also running. SystemX's security analysts did not
know the exact time of the attack. While SystemX's security team had a range of security analysis
and forensic tools to assist in their investigation, they had to analyze events observed from host and
network security monitors: on that day a total of 97,327 events were observed (black vertical bars
in \cref{fig:approach_overview_data}b; some are omitted for simplicity) out of which only 15 events
(shown as colored bars) were related to the attack. The progression of our attack is depicted in 
\cref{fig:approach_overview_data}a. The diamonds, ovals, and rectangles depict IP addresses,
processes, and files, respectively. The attack progressed as follows.

In the \emph{discovery stage}, the attack \emph{repetitively} scanned for vulnerable Apache
Struts portals (i.e., $\epsilon_{1},\dots,\epsilon_4$) in \cref{fig:approach_overview_data}b.
Those scan events did not require immediate action from SystemX's operators because a few scans
do not lead to network congestion, and hence such activities were considered routine. Once a
vulnerable Struts server is identified, the attack gained initial access by exploiting the remote
code execution vulnerability (i.e, $\epsilon_5$) that blends into scan events.

The \emph{gathering stage} in \cref{fig:approach_overview_data}a illustrates each of the attacker's
actions with its corresponding UNIX command. In order to make sure that the system admin was not
present, the attack queried currently active users (i.e., $\epsilon_6$: command \texttt{w}). The
attack obtained the kernel version to prepare for a suitable exploit (event $\epsilon_7$, command
\texttt{uname}). In the \emph{preparation stage}, the attacker established a reverse shell to
his/her machine and transferred the source code of the PE exploit (\texttt{cowroot.c}) and a
sophisticated rootkit (\texttt{knockd.c}) to the legacy memory-mounted disk
(\texttt{/dev/shm}; i.e., $\epsilon_8$) to make sure that no attack traces would be 
presented on the disk file system.

In the \emph{privilege escalation stage}, the attacker obtained a compiler toolchain ($\epsilon_9$: command \texttt{apt}) and compiled ($\epsilon_{10}$: command \texttt{cc}) the
exploit. Unlike Windows environments in which malicious binaries can run directly, SystemX uses a
variety of highly customized Linux distributions for different computing tasks. Thus, attackers
could not deploy binaries directly even if they got the kernel version. In our past
APTs, attackers must compile exploit code for the specific kernel configuration of the target
machine. After executing the compiled PE exploit, the attacker became the superuser: this activity
corresponded to a severe event (i.e., $\epsilon_{11}$). If the attack was not stopped here, the attacker
could compile the rootkit and load (\texttt{insmod}) the rootkit as a kernel module (i.e., 
$\epsilon_{12}$).

In the \emph{persistence stage}, once the kernel module (\texttt{knockd.ko}) has been loaded, the attacker
could maintain \emph{persistent} access by using two rootkit components:
\begin{enumerate*}[label=(\roman*)]
    \item a user-level backdoor that listens at port 9090 to receive
    remote commands, and
    \item a netfilter kernel hook that provides a stealth \emph{port-knocking service} for the
    user-level backdoor, i.e., the port only opens for a short period of time upon receiving a
    specific \quotes{knocking} sequence of three TCP packets such that \texttt{src\_port +
    seq\_number = 1221}.
\end{enumerate*}
Thus, the secret port stays under the radar of network-security scanning tools (which treat these
packets as misrouted or corrupted). Finally, in the \emph{exfiltration} stage, the attacker
extracts secret SSH keys (\texttt{id\_rsa}) and data (\texttt{portal.db}) by using a DNS tunnel
(\texttt{iodine}; i.e., $\epsilon_{13},\dots,\epsilon_{15}$).

\subsection{Motivating Example: The Defense} \label{sec:defenders_perspective}
We describe the defense from the perspective of an experienced security analyst and show how PULSAR
uses machine learning to automate the detection process.

Assuming that an analyst observes a stream of events (see \cref{fig:approach_overview_data}), we
expect that s/he will focus on three key events. First, on observing scans $\epsilon_1-\epsilon_4$, the
analyst might suspect malicious intent, but have low confidence that it might mature into
an attack (a few scans seldom lead to a major attack). Second, the system queries $\epsilon_6,\epsilon_7$ increase the analyst's suspicion. Most benign users do not query detailed system
configurations (although they might). At this stage, the analyst's confidence increases, but not
enough to warrant a reaction. As a result, the decision is to continue to monitor additional
events. The third set of key events is the placement and compilation of the source file in the
memory-mounted disk, i.e., $\epsilon_{8},\epsilon_{9},\epsilon_{10}$. Although these events may be used in legitimate activity to speed up I/O (e.g., an in-memory file system), $\epsilon_{8}$ is
also observed in attacks, as such volatile storage does not leave any forensic evidence after system
reboot. Finally, the analyst observes $\epsilon_{11}$, a \emph{severe} event corresponding to PE. If
the attack is not stopped at this point, the system integrity will be lost, i.e., the attacker will get superuser
permission. However, one must note that PE is not always a malicious event. In some contexts
(especially on personal workstations on enterprise networks) ordinary users might have legitimate
reasons to escalate to a superuser, e.g., to perform an operating system upgrade.

The above scenario assumes there is an experienced analyst who can analyze the events in a stream, presumably with the help of various tools, to make a decision. However, performing such an analysis in real-time
today is generally impossible, as there are an average of 94K events per day. Note that our
hypothetical analyst made his/her decision based on
\begin{enumerate*}[label=(\roman*)]
    \item \emph{repetition} of scans,
    \item \emph{commonality} of the action (since observed previous incidents involved system
    configuration queries), and 
    \item \emph{severe--critical} use of the compilation of the source code on the volatile file system.
\end{enumerate*}
In addition, taken together, the sequence of a \emph{scan}, followed by \emph{system
configuration queries}, and the \emph{placement and compilation} (of source files) in a
memory-mounted disk, is one of the 45 common event sequences in our dataset (see \cref{fig:attack_similarities}).

We present in this paper a machine-learning based approach, based on PGM~\cite{beccuti2015markov,frey1997factor,pearl2011bayesian,sanders1986metasan,halkidi2006resilient,kulkarni2018inability,joshi2011probabilistic}, which builds an FG~\cite{frey1997factor} for each user, and can perform real-time detection. Our approach learns from past events to build
prior knowledge and uses the learned domain knowledge of the system together with observed runtime
events. PULSAR can achieve this in the presence of noisy events, i.e., many users using the system. As
each event is observed, PULSAR adds a factor function (FF) to the event that characterizes its
commonality, severity, or repetitiveness. These factor functions are learned by extracting events
sequences from our dataset of 120 past APTs. At its core, an FF captures how frequently is an event (or
an event sequence) occurred in past APTs and the statistical significance. The FFs together
with observed events, are used to infer attack stages, which allows our decision algorithm
to preemptively stop an attack.

PULSAR processes events as follows. First, each of the scan events $\epsilon_1-\epsilon_4$ has an FF
attached to it to capture the frequency and the significance of repetitive scans. This FF indicates that
the attack has not yet met the stopping threshold (i.e., the attacker is only discovering the
system). Then, the attacker gets a terminal to control the Struts server (event $e_5$). After that, when
PULSAR observes the events \emph{query active users} and \emph{query kernel version} $\epsilon_{6}, \epsilon_{7}$, it adds FFs to the
event to capture these events severity. The FFs indicate that the events are not severe, and thus
the attack continues to be monitored. If any observation of this event was stopped, many legitimate users would be
unable to work. Finally, PULSAR observes the placement and compilation of source code, i.e., $\epsilon_{8}-\epsilon_{10}$, which combined with past events ($\epsilon_6,\epsilon_8,
\epsilon_{10}$) forms a common sequence. While individual events in this sequence can be
legitimate, the whole sequence has occurred frequently in past APTs. Now,
PULSAR outputs the attack stage as preparation (immediately before PE) and notifies analysts to stop the attack.

We quantify PULSAR's preemptive detection capabilities in terms of: i) \emph{preemption before system
integrity violation and before data loss} and ii) \emph{preemption after system integrity violation (SI)
but before data loss (DL)}. These metrics can be quantified in terms of the distance (in hops)
calculated by a function $hop(D_{\sigma},L_{\text{SI+DL}})$ between $D_\sigma$, the stage when an
attack is detected (i.e., \emph{stop} action is suggested), and $L_{\text{SI+DL}}$, the actual
system integrity violation without any data loss. This hop-based metric is time-independent; thus, it
can characterize attacks that happen on various time-scales, e.g., on the order of minutes, days, or
weeks. PULSAR stops the 6-stage APT above (\cref{fig:approach_overview_data}) at its
3\textsuperscript{rd} stage, i.e., at $D_\sigma=\sigma_5$ (preparation), while the system integrity
violation is at $L_{SI+DL}=\sigma_6$ and $hop(D_{\sigma},L_{\text{SI+DL}})=1$.
\cref{fig:approach_overview_data}b shows that PULSAR infers attack stages that approach the
successful completion of the APT. Finally, PULSAR's decision is presented to security analysts.
These decisions could be automated by actuators, e.g., which could redirect malicious network flows~\cite{skowyra2018effective,ujcich2017attain,wang2015floodguard,zhang2014adaptable} or allocate resources for further monitoring~\cite{truta2017predictive}.
\section{Threat Model and Assumptions}
\label{sec:threat_model}

\textbf{Target system.}
This paper considers multiuser, multinode, networked computer systems that provide computational
services in which users remotely execute workloads on internal hosts (physical/virtual machines).
The system is assumed to be benign at the onset. The system may have unpatched vulnerabilities due to the complexity of patching highly
interconnected system components. Indeed, recent surveys~\cite{pashchenko2018esem,lauinger2017thou}
found that 37\% of the top 133k websites still use vulnerable libraries.

We assume that the events from network- and kernel-based monitors are
trustworthy~\cite{garcia2011diversity,beham2013intrusion,babay2018network,liu2018towards,king2005enriching,lee2013high,ma2016protracer,xia2012cfimon,kurmus2011attack,hossain2015towards} and accurate in
capturing attack activities. Since our approach's accuracy depends on monitors, we use an extensive
set of well-configured (e.g., SystemX uses a Bro cluster for network monitoring) and well-protected
monitors (e.g., osquery runs at the kernel-level). While an attacker may tamper with one monitor (on one
host) by using the credential of a local privilege user, it would be challenging to manipulate \emph{all} monitors. We
describe an example attack (\textbf{A2} in \cref{sec:eval}) that manipulates the Bro cluster
to suppress attack-related security events; however, PULSAR still works.

\textbf{Attacker Capabilities.}
This paper assumes that an attacker can pretend to be a legitimate user by using weak/stolen
credentials~\cite{mazurek2013measuring,egele2017towards,barronhoney,han2016shadow} or remote command
execution exploits~\cite{CVE-2017-5638} to compromise internal hosts. PULSAR treats it as a single attack if (1) an attacker moves laterally (e.g., connects by SSH to multiple machines) using the same user account and (2) multiple (coordinated or independent) attackers launch an attack using the same user account. If (1) an attacker moves laterally using different user accounts, or (2) one or more attackers use different entry points and launch attacks using different user accounts, PULSAR treats that as multiple separate attacks.

An attacker may mimic legitimate user activities~\cite{wagner2002mimicry} to obfuscate
attack-related activities. Such an attack only works against TDSs that use a small
sliding window of events (e.g., 9 events in~\cite{phids}) to detect an attack. PULSAR uses a larger
sliding window of 10,000 events, and filters insignificant events (\cref{sec:eval}). If an attacker
fills up PULSAR's sliding window, PULSAR will not be able to detect the attack. However, such
attack activities might cause significant perturbations in system operations, and hence be
observable to the operators.

The boundary of our threat model is that PULSAR cannot preempt an attack if the attacker (e.g., a malicious insider) executes an attack  
\begin{enumerate*}[label=(\roman*)]
    \item in a single step without being persistent,
    \item with no time evolution involving prior events, and
    \item without events in common with any of the past APTs.
\end{enumerate*}
Nonetheless, as we showed in our
results, our threat model can capture a wide variety of attacks, as discussed in
\cref{sec:result}.

\section{Learning and Inference Model} \label{sec:modeling}
This section introduces FGs and demonstrates its advantages in addressing the threat model.

\subsection{Formulating APT Defence as a FG}

\begin{definition}
\label{def:factor_graph}
A \emph{Factor Graph}~\cite{frey1997factor} is a graphical representation of the factor-argument
dependencies of a real valued function. Given the factorization of a function $f(x_1, \ldots,
x_n)$,
\[
    f(x_1, \ldots, x_n) = \frac{1}{Z}\prod_{i = 0}^m{f_{i}(X_i)} \text{, where }  X_i \subseteq 
    \{x_1, \ldots, x_n\}
\]
The corresponding FG $G(X\bigcup F, A)$ is a bipartite graph that consists of variable vertices $X =
\{x_1, \ldots, x_n\}$, factor vertices $F = \{f_1, \ldots, f_m\}$, and arcs $A = \{ (x_i, f_j) \mid
x_i \in X_j\}$. The functions $f_i$ are called \emph{factor functions}.
\end{definition}

\begin{definition}
\label{def:event_timeline}
An \emph{event timeline} is a sequence of events $E_t = \{ e_i~|~i \in
[0,t] \},$ observed at each time step $i$ until the time $t$. We also call such a timeline an
\emph{event stream}.
\end{definition}

\begin{definition}
\label{def:stage_timeline}
A \emph{attack-stage timeline} is a sequence of unknown stages $S_t = \{ s_i~|~i \in [0,t] \}$
corresponding to $E_t$.
\end{definition}

\begin{figure}[!t]
	\centering
	\includegraphics{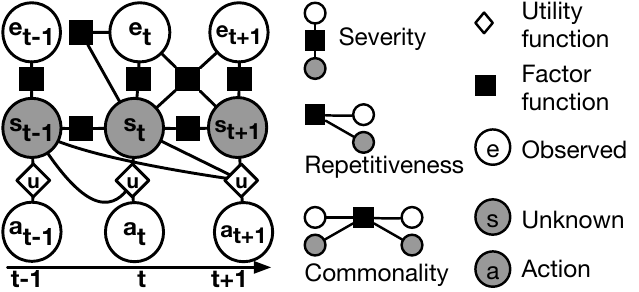}
	\caption{An example Factor Graph at runtime (from \cref{sec:modeling}).}
	\label{fig:factor-graph-and-decision-network}
\end{figure}

PULSAR uses the FG formulation (shown in \cref{fig:factor-graph-and-decision-network}) to describe
the progression of APT by factorizing the joint probability distribution (pdf) of $E_t$ and $S_t$ as
\begin{equation}
    P(E_t, S_t) = \frac{1}{Z}\prod_{c}{f_{c}(X_{c})},
    \label{eqn:pulsar_pdf}
\end{equation}
where $Z$ is a normalization factor that ensures $P$ is probability distribution. Each set of random
variables $X_c$ contains a subset of events $E_c$ and attack stages $S_c$ such that $E_t \cup S_t =
\bigcup\displaylimits_{c} (E_c\cup S_c)$. \cref{sec:factor_functions} will describe how the attack
characteristics (from \cref{sec:sev_rep_com}) of \emph{severity}, \emph{commonality} and
\emph{repetitiveness} can be encoded into FFs and the choice of different $X_c$.
The PULSAR FG makes no assumptions about the causal ordering among events. This is a fundamental strength of PULSAR compared to other techniques like provenance
techniques~\cite{bates2015trustworthy} as it is difficult to define the causal ordering of events in realistic attacks.

\subsection{Formulating \& Training Factor Functions} \label{sec:factor_functions}
The challenges with formulating the FFs for the FG described above (from \cref{eqn:pulsar_pdf}) are 
\begin{enumerate*}[label=(\alph*)]
    \item to find the set $X_c$ by selecting events from $E_t$ and states from $S_t$; and
    \item to find a functional form of $f_c$ that relates the variables in $X_c$.
\end{enumerate*}
PULSAR's definition of FFs presents a three-fold solution to the challenges above, i.e., it uses the
three common characteristics (from \cref{sec:sev_rep_com}) of the attacker actions (behavior) observed
in past attacks as three separate class of FFs. Each attack behavior is represented by a set of
events and its corresponding attack stages as its input (i.e., $E_c$ and $S_c$ above). Although
these variables define a large combinatorial space, PULSAR focuses only on observations that are
\begin{enumerate*}[label=(\alph*)]
    \item highly frequent in a past attack, and
    \item statistically significant, so that the observation reliably indicates an attack.
\end{enumerate*}

We capture the frequency of a sequence of events $E_c$ and attack-stages $S_c$ in a function $q(E_c,
S_c)$ such that
\begin{equation}
    q(E_c, S_c) = 
    \begin{cases}
        P(E_c, S_c), & \text{if } E_c \text{ is obs. in the cur. attack}\\
        0, & \text{otherwise}
    \end{cases}
    \label{eqn:q}
\end{equation}
where $P(E_c, S_c)$ is the probability of observing $E_c$ having states $S_c$ in the training
dataset $\mathcal{D}$. We capture the statistical significance of the event- and
attack-stage--sequence by computing the significance level $p(E_c, S_c)$ (i.e., \emph{p-value}: the
probability that the null-hypothesis is \emph{true}) of observing $E_c$, $S_c$ under the
null-hypothesis that ``the target system is currently not under attack''. To combine both frequent
and significant characteristics, PULSAR uses exponential form FFs, i.e.,
\begin{equation}
    f(E_c, S_c) = \exp \{q(E_c, S_c)(1 - p(E_c, S_c))\}.
    \label{eqn:pulsar_ff}
\end{equation}
The advantages of this exponential form are that
\begin{enumerate*}[label=(\alph*)]
    \item it achieves numerical stability for small values of $q(X_c)$ and $p(X_c)$ (especially 
    when attacks are rare events), and
    \item it ensures that the product of FFs is a convex function.
\end{enumerate*}
Convex-exponential FFs allow us to efficiently optimize the joint pdf in our learning and inference
steps by using the \emph{log-sum-exp} trick. For notational simplification, we use the shorthand
$p_c = p(E_c, S_c)$ and $q_c = q(E_c, S_c)$.

PULSAR maps the problem of learning FFs into finding the parameters $p_c$ using
patterns from past security incidents to capture three characteristics of APTs:
\emph{commonality}, \emph{severity}, and \emph{repetitiveness}. PULSAR utilizes an ensemble of
generic statistical hypothesis testing frameworks~\cite{sheskin2003handbook}, each of which is
customized for the properties mentioned above. Only the FFs which have a high significance ($p_c \leq
0.05$), i.e., that observing the properties under the null hypothesis is unlikely) are considered
for use in \cref{eqn:pulsar_pdf}.

The FFs are trained from an annotated dataset of past attacks $\mathcal{D} = \{ A_i \mapsto [E_t^i,
S_t^i, U_t^i] | \forall i\}$ which consists of a set of attacks $A_i$, each of which includes an
event timeline $E_t^i$, an annotated attack stage timeline $S_t^i$, and an ownership map $U_t^i = \{
u_j ~|~j \in [0,t] ~\land~ u_j \text{ owns } e_j \text{in} E_j^i\}$ which identifies which user is
responsible for which event. In addition, we have a list $\mathcal{U_i}$ which identifies the
malicious users in each scenario. Using this formalism of the input dataset, we consider the
training procedure of each of the FFs presented above as follows.

\textbf{The Severity FF.}
Severity FFs measure the maliciousness of an event. We consider an event malicious when
an attacker has taken control of the system or has caused significant damage. It can be
argued that traditional signature-based TDS (e.g.,~\cite{paxson1999bro}) are a degenerate
(non-probabilistic) case of the severity FF. Hence, in the case of severity FFs, $E_c = \{e\}$ and
$S_c = \{s\}$. To understand the occurrence of an event $e$ in $\mathcal{D}$ we compute two
probabilities $P_{A, s}(e)$, the probability of an event being from attack stage $s$, and $P_B(e)$,
the probability of an event being benign. So
\begin{multline}
    p_{\text{Severity}}(e, s) = \chi^2\left(\left\{
        \frac{P_{A, s}(e)}{\sum_x P_{A, s}(x) + P_B(x)},\right.\right.\\
        \left.\frac{\sum_x P_{A, s}(x) - P_{A, s}(e)}{\sum_x P_{A, s}(x) + P_B(x)}\right\}, 
        \left\{\frac{P_B(e)}{\sum_x P_{A, s}(x) + P_B(x)},\right.\\
        \left.\left.\frac{\sum_x P_{B}(x) - P_{B}(e)}{\sum_x P_{A, s}(x) + P_B(x)}
        \right\}\right),
    \label{eqn:severity}
\end{multline} 
where $\chi^2(\cdot)$ represents the \emph{p-value} of the chi-squared
test~\cite{sheskin2003handbook}. This test is applicable for testing dependencies of categorical
variables, i.e., dependency between the observed events and the attack stage. In the training phase,
the value of $p(e,s)$ for all statistically significant combinations of $e$ and $s$ is computed
offline and stored for use at inference.

\textbf{The Commonality FF.}
Commonality FFs measure the similarity between a sequence of events observed during an attack and
all past known attacks. Thus, conventional similarity measures, such as Hamming distance or string
similarity, are not suitable. To capture the aforementioned characteristics, we quantify similarity
using the longest common subsequence (LCS) measure~\cite{nistlcs,atallah1998algorithms} which outputs
a sequence of events that appear in the same order in two attacks. This LCS approach works with
mimicry attacks in which attackers inject arbitrary noise to the event sequence. Note that attackers
may not simply change the order of events to trick our approach, because many events depend on each
other, e.g., executing the binary file of a memory exploit first requires the successful compilation
of the exploit from the source code.

In the training phase, the LCS $L_{i,j}$ is computed for all pairs $(A_i, A_j)$ (such that $i\neq
j$) of attacks in $\mathcal{D}$. Each unique value of $L_{i,j}$ has an annotated attack stage
$S_{i,j}$. The commonality FF's $p_c$ is calculated as
\begin{multline}
    p_{\text{Common}}(L_{i,j}, S_{i,j}) = \mathbb{F}\left(\left\{
        \frac{P_{A, S_{i,j}}(L_{i,j})}{\sum_x P_{A, S_{i,j}}(x) + P_B(x)},\right.\right.\\
        \left.\frac{\sum_x P_{A, S_{i,j}}(x) - P_{A, S_{i,j}}(L_{i,j})}{\sum_x P_{A, S_{i,j}}(x) + P_B(x)}\right\},
        \left\{\frac{P_B(L_{i,j})}{\sum_x P_{A, S_{i,j}}(x) + P_B(x)},\right.\\
        \left.\left.\frac{\sum_x P_{B}(x) - P_{B}(L_{i,j})}{\sum_x P_{A, S_{i,j}}(x) + P_B(x)}
        \right\}\right),
    \label{eqn:commonality}
\end{multline}
where $\mathbb{F}(\cdot)$ is the \emph{p-value} of the Fisher exact test~\cite{sheskin2003handbook}
and $P_{A, S_{i,j}}$ and $P_B$ are defined as above. We test for significance of the LCS by using
this test on a small sample size, e.g., a limited observation of attacks, as malicious
attacks are rare compared to legitimate activities in our data source. All statistically significant
combinations of $L_{i,j}$, $S_{i, j}$ and $p(L_{i,j}, S_{i,j})$ are stored for use at inference.

\textbf{The Repetitiveness FF.}
A repetitiveness FF measures the periodicity of an event. Regular events occurring in a regular period
are often results of the repeated execution~\cite{kwon2016catching} of automated scripts and hence
can be used as an indicator of automated attacks at an early stage (e.g., periodic port scan
events~\cite{doupe2012enemy,schagen2018towards}).

For every event $e$ which is annotated with attack stage $s$ in an attack $A_i$, PULSAR computes a
the frequency $H_{s, e, n, k}^{(i)}$ of occurrence (i.e., number of repetitions) of $(e, s)$ in the
time interval $[nk, (n+1)k]$ during the attack $A_i$. Then
\begin{equation}
    p_{\text{Repetitive}}(e, s) = \min_{i, n, k} \text{DW}\left(\left\{H_{s, e, n, k}^{(i)} | A_i \in \mathcal{D} \right\}\right),
\end{equation}
where $\text{DW}(\cdot)$ is the \emph{p-value} of the Durbin-Watson test~\cite{sheskin2003handbook}. These
$\text{DW}$ tests are computed in training. The statistically significant repetitive events are
stored for use in inference.

All three FFs above are generated by statistical tests based on the counts of past events. Thus, when a new attack is observed, PULSAR does not need to re-learn the FFs. Instead, it increments the frequency of the events sequences (according to the events in the new attack) and reruns the statistical tests above. This notion of incremental training proposition is useful for systems that observe new attacks on a daily basis. For attacks that introduce new events, we discuss them in \cref{sec:discussion}.

\textbf{Constructing FGs.} At runtime, the FG only contains an event $e_t$ and an unknown stage
$s_t$. PULSAR selects FFs (from the learned FFs above) that are statistically significant and adds
these FF to the FG to connect $e_t$ and $s_t$. When a new event $e_{t+1}$ is observed, the
above procedure is repeated. In addition, PULSAR adds a transition FF: $f(s_{t}, s_{t+1})$ that
captures the probability that a stage $s_t$ leads to a stage $s_{t+1}$. This probability is represented by a $11\times11$ matrix. Finally, PULSAR performs
inference on unknown attack stages.

\subsection{Inference On Factor Graphs} \label{sec:decision}

PULSAR uses the constructed FG above and its inference algorithm (see
Algorithm~\ref{alg:inference}) to predict the sequence of unknown attack stages $S_t$ associated
with a sequence of observed events $E$ up to a time $t$. Based on how an analyst responds to an
attack at SystemX, we consider a three-tiered decision model (see \cref{tab:actions}) that separates
actions from their implementations, which might be system dependent. Based on that estimate of
$S_t$, a preemptive action $a_t$ (e.g., \emph{no-op}eration, \emph{monitor} a user closely--via
deep packet inspection, or \emph{stop} the user immediately) is suggested to an analyst.

\begin{algorithm}[!t]
	\small
	\SetAlgoLined
	\SetKwInOut{Input}{Input}\SetKwInOut{Output}{Output}
	\Input{Set of learned factor functions $F$, \\
		An event timeline $E_t$ up to a time $t$}
	\Output{Sequence of attack stages $\widehat{s}_t$, \\
		Action $\widehat{a}_t \in \mathcal{A}$}
	FG $\gets \varnothing$\\
	$\Delta$ $\gets 0.05$\\
	\For{$e_i \in E_t$}{
		add an event node $e_i$ to the FG \\
		add an attack stage node $s_i$ to the FG \\
		add a transition FF between $s_i$ and $s_{i-1}$ \\
		\For{$f \in F$}{
			\If(\tcp*[f]{select a ff}){$q_f > 0$ and $p_f \leq \Delta)$}{
				add the factor function $f$ to the FG \\
			}
		}
	} 
	$\widehat{s}_t \gets TRW(P(E_t,S_t)$\tcp*[f]{using Tree Reweighted to find $argmax{s}_t$} \\
	$\widehat{a}_t \gets \argmax_{a}\sum_{i \in S} P(\widehat{s}_t = i)\times u(a,i)$ \\
	\Return{$(\widehat{s}_t, \widehat{a}_t)$}
	\caption{Construction and inference on FG}
	\label{alg:inference}
\end{algorithm}

PULSAR's inference procedure finds the most probable attack stages that minimize the energy~\cite{yedidia2005constructing} of the
FGs. Since our FG can have loops (two variables may
be connected by more than one factor function), loopy BP does not guarantee convergence: the
inferred attack stages may reach the global minimum energy of the FG. Therefore, we adopted the sequential tree-reweighted (TRW) message passing
scheme~\cite{kolmogorov2006convergent} that: i) decomposes the loopy FG into smaller FGs without a
loop,  ii) guarantees that the inference procedure will converge. Using TRW, PULSAR infers a maximum
likelihood estimate (MLE) for a sequence of \emph{unknown} attack stages ${\widehat{S}}_{t}$ (up to
time $t$) associated with an event timeline $E_{t}$. Further, it outputs a confidence level for each
stage, as an event is observed. That means finding ${\widehat{S}}_{t} = \argmax_{S}{P\left( E_t,S
\right)}$, where $P\left( E_{t},S \right)$ is defined using the factorization
equation (\cref{sec:modeling}) in FGs. The result of this optimization procedure is the most likely
sequence of attack stages associated with observed events. This procedure provides the step-by-step estimate of attack stages for analysts.

\begin{table}[!t]
    \centering
    \rowcolors{2}{gray!15}{white}
    \caption{List of preemptive actions in the PULSAR reward model (see \cref{fig:rewards}).}
    \begin{tabular}{cp{7cm}}
        \toprule
        \textbf{Action} & \textbf{Description} \\
        \midrule
        $\alpha_1$& Benign user behavior, no immediate action needs to be performed. \\
        $\alpha_2$& An attack may be in progress, and further monitoring is required. \\
        $\alpha_3$& An attack is imminent and must be stopped immediately. \\
        \bottomrule
    \end{tabular}
    \label{tab:actions}
\end{table}

\subsection{Decision on maliciousness of an attack}\label{sec:decision_model}
The FG model is extended with ``reward factor functions'' to produce a decision model that
determines an optimal action based on the inferred attack stage (see
\cref{fig:factor-graph-and-decision-network}) to support learning~\cite{erdt2015evaluating}. The
decision model has two new components:
\begin{enumerate*}
    \item a finite action space $\mathcal{A} = \{\alpha_1,\alpha_2,\alpha_3\}$ (see 
    \cref{tab:actions}), and
    \item a reward function $u:\mathcal{A} \times S\rightarrow \mathbb{R}$.
\end{enumerate*}
The output of the decision model
allows security analysts to closely track the progress of an ongoing attack and implement response
strategies. The decision model is useful when PULSAR is not confident in its inferred attack
stage (i.e., the prediction probability is close to 50\%). In this case, a \emph{monitor}
action ($\alpha_2$) is recommended instead of immediately stopping user activity, which would
otherwise lead to FPs. At time $t$, the decision model uses the inferred attack stage $\widehat{s}_t$, and
its probability $P(\widehat{s}_t)$, to choose an action $\widehat{a_t}$ that maximizes the reward
over $P(\widehat{s}_t)$ in~\cref{sec:decision_model}.

\begin{figure}[!t]
    \centering
    \includegraphics[width=\columnwidth]{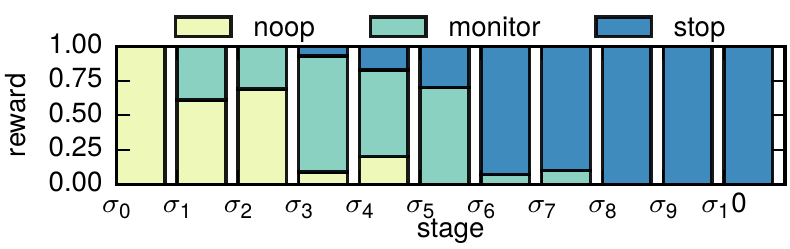}
    \caption{Reward function $u(a,s)$ for each attack stage.}
    \label{fig:rewards}
\end{figure}

\begin{algorithm}[!t]
    \small
    \SetAlgoLined
    \SetKwInOut{Input}{Input}\SetKwInOut{Output}{Output}
    \Input{Training data $D$ of size $n\times11$, \\
        Priors $\mu_p = \{\mu_{a} | a \in \mathcal{A}\}$, and
        $\Sigma_p = \{\Sigma_{a} | a \in \mathcal{A}\}$}
    \Output{Reward function $u(a,s)$}
    \BlankLine
    \tcc{Definition: $\text{GMM}(x | k, \mu, \Sigma) = \sum_{i = 1}^{k} w_i \times \mathcal{N}(x | 
        \mu_i, \Sigma_i)$}
    \BlankLine
    $w, \mu, \Sigma \gets \text{EM}(D, k, \mu_p, \Sigma_p)$\tcp*[f]{Train $k = 3$ GMM model}\\
    $u \gets \varnothing$\\
    \For(\tcp*[f]{Corresponding to $\{\alpha_1, \alpha_2, \alpha_3$\}}){$a \in \mathcal{A}$}{
        \For(\tcp*[f]{Corresponding to $\{\sigma_0, \ldots, \sigma_{10}$\}}){$s \in [0, 10]$}{
            $u(a,s) \gets w_a \times \mathcal{N}(I_{11}[s,:] |~\mu_a, \Sigma_a) $\\
        }
    }     
    \Return{u}
    \caption{Computing the reward function $u(a,s)$.}
    \label{alg:gmm}
\end{algorithm}

To model a mapping from attack stages (obtained from FG's inference) to three actions that analysts
at SystemX do, we use a 3-component Gaussian Mixture Model (GMM; corresponding to the 3 actions
described in \cref{tab:actions}) with priors to construct the reward functions. Gaussian has a
conjugate prior distribution that augments these functions with operational information about past
incidents, i.e., actions that SystemX's security team has taken in past incidents. A reward function
$u(a,s)$ (see \cref{fig:rewards}) on an action $a$ being taken at stage $s$ is estimated over a 11-D
space (corresponding the attack stages $\sigma_0, \ldots, \sigma_{10}$). Each of the components
(clusters) in the GMM corresponds to one of the actions $\{\alpha_1,\alpha_2,\alpha_3\}$.  For
example, in the early attack stages $\sigma_0,\sigma_1$, the no-op action $\alpha_1$ is preferred.
In the intermediate stages $\sigma_2,\sigma_3$, the monitor action $\alpha_2$ (i.e., deep packet
inspection) is preferred. Finally, in attack stages $\sigma_4,\ldots,\sigma_{10}$, the stop action
$\alpha_3$ has a higher reward than $\alpha_1$ \& $\alpha_2$. The training procedure (using
\emph{expectation maximization}) is described in \cref{sec:decision_model}.
\section{Evaluation Setup} \label{sec:eval}
Real-world APTs blend their attack events with background activities of legitimate users running production workload. The question is whether PULSAR can identify real APTs in a timely manner. We demonstrate PULSAR's accuracy and performance via three experiments. \emph{First}, we evaluated PULSAR on 120 past APTs retrospectively. \emph{Second}, we evaluated PULSAR's detectability on ten unseen attacks constructed using the IBM Threat Intelligence Index and injected in live traffic where normal security events were intermingled with the unseen attack events. \emph{Third}, PULSAR was integrated into SystemX's security infrastructure, for a month-long trial, while measuring its performance overhead.

\begin{figure}[!t]
    \centering
    \includegraphics[width=\columnwidth]{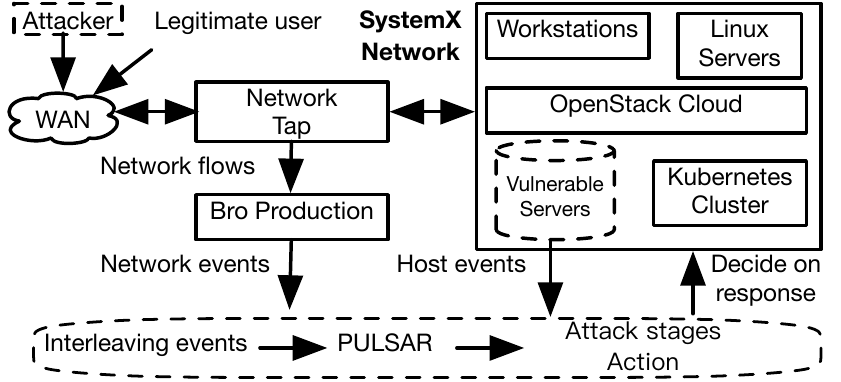}
    \caption{Testbest architecture for evaluating PULSAR.}
    \label{fig:testbed}
\end{figure}

\begin{table}[!t]
    \centering
    \small
    \caption{Summary of dataset used in our experiments.}
    \rowcolors{2}{gray!15}{white}
    \begin{tabular}{llll}
        \toprule
        \textbf{Description} & \textbf{Past} & \textbf{Unseen} & \textbf{Production} \\
        \midrule
        Total events & 235K & 1.25M & 4.9M\\
        Unique events & 105 & 68 & 42 \\
        Number of APTs & 120 & 10 & -- \\
        Type of APTs & Known & Unseen & Attempted\\
        Data duration & 10 years & 1 week & 1 month \\
        \bottomrule
    \end{tabular}
    \label{tab:dataset}
\end{table}

\textbf{Experiment 1: Evaluation based on successful past APTs.} 
To provide a baseline estimation of the accuracy of PULSAR in correctly detecting past APTs, data on 120 past APTs is divided
into two disjoint sets: (i) a set of 60 attacks (2008--2009)
for training PULSAR’s factor functions and (ii) a set of 60 attacks (2010--present) for evaluating PULSAR’s accuracy in detecting the past APTs\footnote{SystemX has observed fewer attacks in recent years than before 2010 because of improved security policies and authentication systems.}.

\textbf{Experiment 2: Evaluation based on APTs unseen at SystemX.} To test the efficacy of PULSAR in detecting attacks never observed in SystemX, we reconstructed and replayed ten attack scenarios (described in \cref{tab:scenarios}). For this, we studied the mechanisms of high-profile attacks~\cite{fbbreach,aptnotes} (e.g., the Equifax data breach~\cite{equifax}) and reconstructed attack scenarios using top ten attack techniques from the IBM Threat Intelligence Index~\cite{ibmxforce17}. These attack scenarios use different techniques, such as local privilege escalation in \textbf{A2, A8} and remote exploit in \textbf{A1, A6}; a custom rootkit in \textbf{A1}; a custom backdoor in \textbf{A4}; and custom exploits in \textbf{A9, A10}. None of these attacks have been observed in the past at SystemX. 

The attack scenarios in \cref{tab:scenarios} represent major data breaches~\cite{equifax}, credential stealing~\cite{fbbreach}, and system integrity violations (e.g., escaping from the isolation mechanism of Linux containers). 
Thus, they show both the depth (in terms of sophistication) and the breath (in terms of the variety of techniques) of APTs. The attack scenarios were in live traffic during production workloads at SystemX; thus, it is not possible to precisely characterize the similarity between the scenarios and past APTs.

\begin{figure}[!t]
	\centering
	\includegraphics[width=\columnwidth]{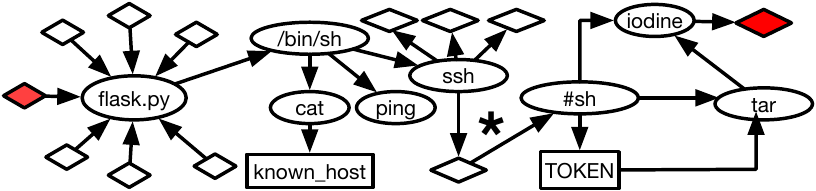}
	\caption{An example sophisticated scenario (\textbf{A2}) that uses a coordinated DDoS attack and an OAuth bypass exploit.}
	\label{fig:a2}
\end{figure}

Because of space constraints, we describe just one sophisticated attack scenario (\textbf{A2}) that attempts to manipulate security monitors. In \textbf{A2}, the attacker ran a mimicry attack against a Python \texttt{flask} server that authenticates users using a vulnerable \texttt{pysaml2} library (\cref{fig:a2}). To mislead security analysts and blend the real exploit into legitimate activities, the attacker attempted to flood Bro by launching hundreds of legitimate-looking requests to the \texttt{flask} server and other servers. In this case, Bro did not have a signature of the exploit (OAuth bypass~\cite{CVE-2017-1000433}), and thus, allowed the attacker to access the vulnerable \texttt{flask} server. PULSAR, however, received the following events from Bro and the host monitors: i) intermittent packet capture losses while production workloads were not high; ii) accessing a list of internal hosts (\texttt{known\_host} file); and iii) unscheduled internal network scans. Combining above events, PULSAR recognized and stopped this attack at the lateral movement ($\sigma_7$) in which the attacker was attempting to access other hosts.

Accurate reconstruction and replay of such attack scenarios, however, presents several challenges.

\begin{table*}[!t]
    	\footnotesize
	\centering
	\begin{varwidth}[b]{\textwidth}
		\centering
		\caption{Ten representative APTs that use top-10 and representative attack vectors from the IBM
			Threat Intelligence Index~\cite{ibmxforce17}.}
		\rowcolors{2}{gray!15}{white}
		\resizebox{\textwidth}{!}{%
			\begin{tabular}{p{0.2cm}p{1.5cm}p{8cm}p{4.5cm}}
				\toprule
				\textbf{ID}& \textbf{Name} & \textbf{Description} & \textbf{Common events with past attacks} \\
				\midrule
				\textbf{A1} & Port knocking &
				Exploit a remote code execution vulnerability CVE-2017-5638 in an Apache Struts web server and read a database of personal information.  & Network scan, get kernel version, access mem disk, compile, new kernel module.
				\\
				\textbf{A2} & DDoS and OAuth&
				Flood network monitors with a high-volume stream of benign-looking requests, forcing the monitor to drop packets. Concurrently, attackers exploit an auth. bypass bug CVE-2017-1000433 in the \texttt{pysaml2} library to steal access tokens. & High network flows, packet loss, access \texttt{known\_host} file, scan internal servers, excessive POST req.
				\\
				\textbf{A3} & Container Escape &
				Exploit a heap vulnerability CVE-2017-5123 on a Kubernetes cluster to read SSH keys in hosted containers. & Get kernel version, compilation, access \texttt{id\_rsa} file.
				\\
				\textbf{A4} & SSH Keylogger &
				Deploy an SSH keylogger into an OpenSSH server, to record credentials of subsequent user logins. & Concurrent login, compilation, restart system service, unknown SSH client.
				\\
				\textbf{A5} & Ransomware &
				Spread ransomware that uses CVE-2017-0144. The ransomware registers itself as a system service to scan for and steal secret files. & New sys. service, scan internal SMB servers, transfer .exe files via SMB.
				\\
				\textbf{A6} & Shellshock &
				Exploit CVE-2014-6271 and setup DNS tunnels to exfiltrate stolen data. & packet loss, get kernel version, access \texttt{/etc/password}, excessive DNS reqs.
				\\
				\textbf{A7} & CPU Bug &
				Exploit a local privilege escalation vulnerability CVE-2017-5754 to steal source code files and extract passwords in the Firefox browser~\cite{lipp2018meltdown}. & Get kernel version, get source code file, compile, excessive POST requests.
				\\
				\textbf{A8} & VM Escape &
				Exploit a Virtual Box vulnerability CVE-2018-2676 to control the host machine and extract private GPG keys. & Access mem disk, compile, access \texttt{secring.gpg}, excessive UDP requests.
				\\
				\textbf{A9} & Race Condition &
				Exploit a race condition in a digital wallet platform to steal the private key. & High network flows, access \texttt{privkey.pem}, excessive ICMP requests.
				\\
				\textbf{A10} & Obj Deserialization &
				Exploit an object deserialization vulnerability in the Python pickle module to steal private TLS keys. & Sensitive commands in HTTP request, access \texttt{privkey.pem}.
				\\ \bottomrule
			\end{tabular}%
		}
		\label{tab:scenarios}
	\end{varwidth}%
\end{table*}

\emph{Challenge 1.} The emulated attack scenarios must include background traffic (i.e., legitimate user activities and background noise such as legitimate port scans) in order to be realistic. \cref{fig:testbed} shows the experimental setup used to collect both types of traffic. Each attack scenario was setup in a vulnerable server (i.e., one that contained software with a known vulnerability) alongside SystemX's workstations, servers, an OpenStack cloud, and a Kubernetes cluster. We simulated a remote attacker accessing the system through the WAN connection to exploit the vulnerable servers in the attack scenarios. Network flows in/out of SystemX (with an average of 12Gbps and a maximum of 40Gbps) were captured via a network tap device at the network border by a Bro cluster of 16 nodes to output network events. Host logs were captured via host-auditing tools and collected via \texttt{rsyslog} to output host events. Both host and network events were collected for preprocessing and analysis by PULSAR. The events were filtered into logical channels. Each channel was a stream of events that was associated with a user. PULSAR subscribed to each channel and constructed a per-user FG, and inferred attack stages/decisions at runtime.

\emph{Challenge 2.} The volume of events generated by monitoring tools in the production system is quite significant and it is possible for attack events to be hidden in this large volume of events. \cref{fig:alerts-per-month} shows that the
test environment experiences a daily average of 94,238 events for a total of 2,032 registered users (at peak times: no more than half of the users use the system concurrently). Although PULSAR processes events as they arrive, it filters out insignificant events by only adding statistically significant (identified by a low p-value) factor functions. Such filtering significantly reduces the number of events in both our training and testing data. PULSAR filtered this stream of an average of 94,238 events to an average of 1,885 significant events that it performed inference. Thus, PULSAR is robust to spurious or noisy events, e.g., ones triggered by benign user activities.

\emph{Challenge 3.} Malware or malicious traffic can proliferate out of control~\cite{gao2017containerleaks}. To ensure full control over the replayed attack scenarios, we contained each attack in a Linux container and 
further encapsulated it in a virtual machine with limited capabilities. All containers used in our
experiments ran in a network sandbox that implemented a Layer-3 private overlay network on a
separated Classless Inter-Domain Routing block.

\emph{Challenge 4.} Reproducing old, vulnerabilities is challenging because new Linux distributions already patch old ones out of the box. For example, to reproduce the Container Escape vulnerability, one would have to obtain not only an old Linux distribution released just before the vulnerability announcement date, but also all dependent package repositories. To address this issue, we built a tool to create old Linux containers at any point in the past (2005--present) using the Debian Linux snapshot repository. This tool allows us to reproduce network-based attack scenarios that include i) a vulnerable server container with an unpatched kernel, user applications, and their dependencies; and ii) an attack container with corresponding exploits (e.g., ICMP tunneling tools). Thus, our approach is more extensible than Metasploit, which only provides exploit code, or a tool introduced in~\cite{mu2018understanding}, which only provides scenarios for memory error vulnerabilities on the host.

With the above challenges solved, we created a testbed environment for automated replay of APTs in production. The data collected from the replayed APTs were used to measure the accuracy and preemption capabilities of PULSAR.

\emph{Accuracy measurement.} We compared PULSAR's accuracy in detecting the ten unseen APTs with that of the \emph{event from any monitor (EA)}, \emph{event correlation (EC)}, and \emph{event statistical anomaly (ES)} techniques
currently deployed at SystemX. The \emph{EA} method is a degenerate approach in which every event is declared to be an attack. The \emph{EC} is implemented using Bro and osquery~\cite{broosquery}; it relies on i) an extensive signature database aggregated from more than 30 anti-virus vendors~\cite{teamcymru}) and ii) manually created correlation rules, e.g., a brute-force SSH login followed by a successful \texttt{root} login, based on host/network event types (similar to the alert correlation technique in~\cite{valeur2004comprehensive}). The \emph{ES} includes a set of statistical measures (similar to ones in~\cite{freeman2016you,ngai2007efficient}) calculated from the outputs of
the deployed security monitors, e.g., days since last login.

\textbf{Experiment 3: Performance and accuracy of PULSAR in production at SystemX.}
To validate PULSAR’s runtime performance in the production network while
real workloads were running, we deployed PULSAR in SystemX for one month and measured its memory consumption, event processing latency, and observed its decisions.

To determine the performance impact of PULSAR on the regular workload, we measured i) runtime memory used by PULSAR, and ii) the latency of PULSAR's algorithms (implemented as a single but multithreaded process based on a C++ OpenGM framework~\cite{andres2012opengm}). All measurements were run in a single 56-core 2 GHz Intel Xeon E5 with 128GB RAM. 

\section{Results}
\label{sec:result}
This section presents the results of the experiments above.

\begin{figure}[!t]
    \vspace{-2ex}
    \centering
    \includegraphics[width=\columnwidth]{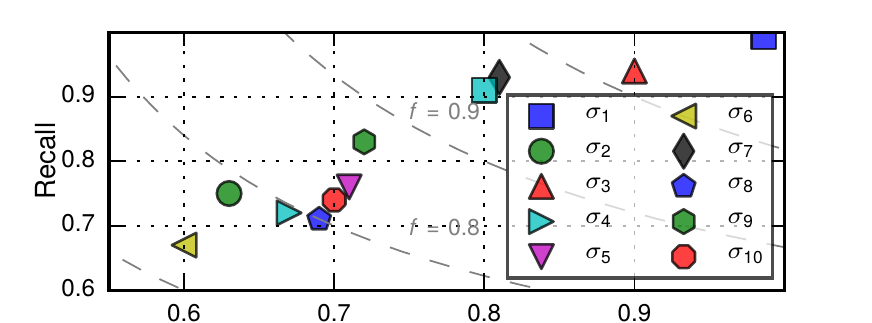}
    \caption{Attack stage detection accuracy in past APTs. Dotted contour lines represent the \emph{F-measure} of the precision-recall spectra. The majority of
        detected attack stages are clustered close to \emph{F-measure}$=1$, i.e., better
        precision-recall.}
    \label{fig:precision_recall}
\end{figure}

\textbf{Result 1: Evaluation based on data on successful past APTs.}
\cref{fig:precision_recall} shows PULSAR's attack stage detection accuracy in terms of the \emph{F-measure} of the precision-recall spectra. F-measure is the 2-times harmonic mean of precision and recall, is upper-bounded by 1, and maps directly to the
effectiveness of PULSAR, with larger numbers signifying better
accuracy. The F-measure range is $[63.3\%,98.9\%]$ with an average accuracy of attack stage detection of 93.5\%. PULSAR shows lower effectiveness in accurately identifying attack stages corresponding to the attack preparation phase ($\sigma_6$ was accurately detected only 63.3\% of the time). The reason for the lower accuracy is 
that many events (e.g., download) generated at the preparation stage overlap with legitimate user activities. For example, a code download frequently corresponds to a legitimate user activity). At SystemX, we have observed
that 89\% of users who perform file downloads are legitimate.

Even so, PULSAR detected 55 out of 60 attacks and performed the best
in detecting the benign stage $\sigma_0$ (which it did 98.9\% of the time with a low FPR) and the gathering stage $\sigma_3$ (which it did 92.0\% of the time -- showing that it could preempt attacks before system integrity violation). 

\begin{figure}[!t]
    \centering
    \includegraphics[width=\columnwidth]{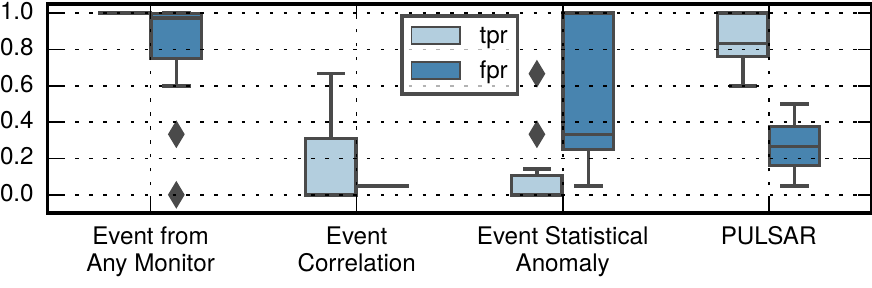}
    \caption{TPR and FPR of the techniques 
    across attack stages. The boxes represent quartiles; the notch represents the median, and the 
    whiskers represent the maximum and minimum.}
    \label{fig:comparison}
\end{figure}

\textbf{Result 2: Evaluation based on data on ten unseen attacks injected in live traffic at SystemX.}
PULSAR’s accuracy in detecting attack stages in the ten unseen attacks is quantified in \cref{fig:comparison} in comparison with the three methods (described in \cref{sec:eval}) currently deployed in SystemX. First, we studied the true-positive rate (TPR), which is
representative of the overall accuracy of the PULSAR system in correctly identifying APTs. 
\cref{fig:comparison} shows that PULSAR significantly outperforms the EC and ES techniques (\cref{sec:eval}). The TPR for PULSAR is 84.8\%, while it is 18.4\% and 11.4\% for EC and ES, respectively. The EA has 100\% accuracy because it indicates an APT for any observed event. Second, we studied the false-positive rate (FPR) which was representative of the number of events generated by legitimate users but detected as attacks. The FPR is important from a security administrator perspective, as a high
FPR can overwhelm a human operator’s ability to react to
an attack quickly. We observe (from \cref{fig:comparison})
that the EC offers a near-zero FPR (which is expected, because EC matches specific event patterns) and that the ES performs poorly on the FPR (the EA has a 99.7\% FPR because all events, including ones from legitimate users, are considered attacks). PULSAR’s small but non-zero FPR (0.02\%) can be attributed to production traffic that interleaves with attack-related activities when we replay a given attack. This traffic adds noise to the stream of events observed by PULSAR. In the context of TPR and FPR, PULSAR offers a trade-off between high attack detection accuracy and a low FPR.

PULSAR preempts attacks and does not need to be informed by monitors of severe events (at the end of attacks). \cref{tab:preemptiveness} summarizes the effectiveness of the four methods (compared in this analysis) in terms of two metrics: (i) preemption of system integrity violation and data loss (SI+DL) and (ii) preemption of data loss while allowing integrity violation (DL). The third column in \cref{tab:preemptiveness} shows the median number of stages (hops) that each method is able to detect the attack before system integrity violation. The last column in \cref{tab:preemptiveness} shows the FPR when we evaluated PULSAR on ten unseen scenarios ran in live production traffic that produced a total of 1.25M events (\cref{tab:dataset}). The \emph{EC} has a low FPR (0.051\%), however, it preempts only one attack out of ten before system integrity violation. Although PULSAR could not stop all system integrity violations, it was able to stop eight attacks out of ten at least one hop before system integrity violation without needing to observe of any critical event (e.g., installation of a kernel module) from underlying monitors. In addition, PULSAR stopped all ten attacks up to three stages (in the median case) before data loss. PULSAR has the lowest FPR (i.e., fraction of events incorrectly classified as attacks), 0.020\%, which is $2\times$ better than that of the second-best method (i.e., 0.040\% in ES).

\textbf{Result 3: Evaluation based on data from PULSAR deployment in the production SystemX.}
A key performance metric for characterizing PULSAR deployment in the production environment is
the observed FPR, in terms of both the percentage and the
absolute number of false alerts~\cite{axelsson1999base}. Given increasing
amounts of network traffic and host activities, even a small FPR can quickly overwhelm a security
analyst's ability to react to an attack. We tested PULSAR in a production environment in which attack traffic was naturally interleaved with legitimate user activities, i.e.,
background network traffic and host activities, making detection significantly more difficult. \cref{tab:alerts} shows the list of the top 5 most/least frequent events. Even under those noisy operational conditions, PULSAR had an impressive FPR of 0.009\%, which corresponds to an average of nine false detections out of an average of 94K daily events. Three out of nine false detections were correlated with three known malicious downloads (shown in \cref{tab:alerts}), in which malicious files (e.g., discussed in~\cite{rahbarinia2017exploring,stock2016kizzle})  were downloaded because of user mistakes, but none of those malicious files was executed. The remaining six out of nine false detections were related to Apache Struts exploit attempts followed by unusual host activities (e.g., disabling of Bash history logging).

\begin{table}[!t]
    \centering
    \small
    \caption{Summary of an early detection result for ten attacks}
    \rowcolors{2}{gray!15}{white}
	\resizebox{\columnwidth}{!}{%
    \begin{tabular}{lllll} 
        \toprule
        \textbf{Method} & \textbf{Median hop} & \textbf{SI+DL}  & \textbf{DL} & \textbf{False Positive} \\
        \midrule
        EA & 4 & 10 & 10 & 99.706\% \\
        EC & 1 & 1 & 2 & 00.051\% \\
        ES & 0 & 0 & 3 & 00.040\% \\
        \textbf{PULSAR} & 3 & 8  & 10 & 00.020\%\\
        \bottomrule
    \end{tabular}%
    }
        \begin{tablenotes}
        \item \textbf{SI}: System integrity violation; \textbf{DL}: Data loss
        \item \textbf{SI+DL}: Attacks stopped before SI and DL
        \item \textbf{DL}: Attacks stopped after SI but before DL
        \end{tablenotes}
        \label{tab:preemptiveness}
\end{table}

\begin{table}[!t]
    \centering
    \small
    \caption{Listing of top 5 most frequent and least frequent alerts during a one-month deployment in production (2018).}
    \rowcolors{2}{gray!15}{white}
    \resizebox{\columnwidth}{!}{%
        \begin{tabular}{llp{2cm}p{4cm}}
            \toprule
            \textbf{Count} & \textbf{Service} & \textbf{Name} & \textbf{Description/Example} \\
            \midrule
            $145.9$K& SSH & Subnet\_Scanner& Scan for SSH hosts\\
            $133.8$K& SSL & Invalid\_Cert& Invalid server certificate\\
            $23.4$K & HTTP & Struts & Exploit Apache Struts\\
            $1.9$K & DNS & Excessive & Large outgoing DNS requests\\
            $1.8$K & HTTP & Shellshock & Attempt to exploit Bash\\
            \midrule
            $175$ & RDP & Brute Force& Remote desktop login\\
            $93$ & ARP & Unknown\_Host& New host on internal network\\
            $39$ & HTTP & Exposed & An internal server is exposed\\
            $36$ & HTTP & SQL\_Injection& Inject SQL commands\\
            $3$ & HTTP & Bad\_Download & A known malicious file\\
            \bottomrule
        \end{tabular}%
    }
    \label{tab:alerts}
\end{table}

\begin{figure}[!t]
    \centering
    \includegraphics[width=\columnwidth]{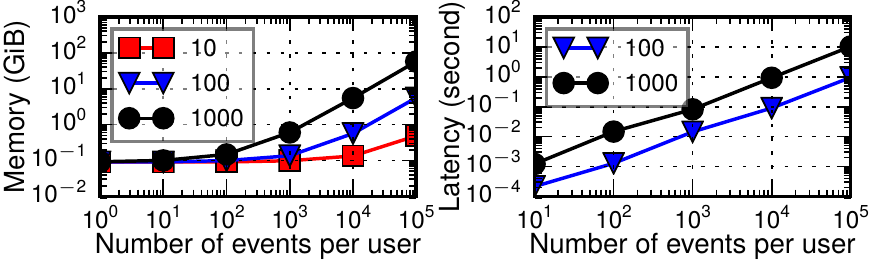}
    \caption{Memory and latency performance of PULSAR for 10,100,1000 users with a varying size of window of events.}
    \label{fig:latency}
    \vspace{-3px}
\end{figure}

\emph{Runtime Performance.} Since SystemX has hundreds of active users a day, we configured PULSAR to handle at most 1,000 concurrent users with a sliding window size of 10,000 events per user. The window size must be long enough to accommodate long-duration APTs~\cite{aptnotes}. In this configuration, PULSAR can handle events for APTs that last up to 217 days (\nicefrac{10,000 events}{46 events per day for a user}), assuming 46 events per user per day as calculated in\cref{sec:eval}. PULSAR takes a median of 1.06 seconds (variance = 0.03) from observing an event to making a decision (\cref{fig:latency}), which is well within the 31-minute inter-event arrival rate of events for each user in SystemX (as calculated in \cref{sec:eval}). PULSAR requires only 126MiB of memory for each monitored user (i.e., 126GiB of memory for the production deployment of 1,000 users; see \cref{fig:latency}).

The PULSAR algorithm's scales as $O(n)$, where $n$ the number of nodes (which includes the numbers of observed events, unknown states, and actions) and the number of FFs (which is fixed from the training data and does not change at runtime) in the graph~\cite{frey1997factor}. PULSAR is currently implemented as a single process, and this limits the number of users that PULSAR can handle at the same time (1000). PULSAR can leverage~\cite{zaharia2016apache} to scale to a larger number of users.
\section{Discussion}
\label{sec:discussion}

\textbf{Explaining PULSAR's success.} As we pointed out in our threat model (\cref{sec:threat_model}), the success of our approach depends on some overlaps between unseen APTs and past APTs. As we have shown in our results (\cref{sec:result}), PULSAR has learned prior knowledge (in terms of commonality, severity, and repetitiveness FFs) from a broad set of 120 past APTs. In fact, each unseen attack (in our ten scenarios shown in \cref{tab:scenarios}) has an average of six key events (not counting excessive scan events) out of a maximum of 30 events and a minimum of 12 events that are common with past APTs. This relatively modest level of similarity with past events enables PULSAR to detect unseen attacks. It is important to note that the training data had only 
thousands of events, while PULSAR in production observed millions of background events. This means the number of attacks (albeit many unsuccessful) has increased significantly.

\textbf{Generalizing PULSAR trained knowledge to other systems.} A pre-trained instance of PULSAR (e.g., based on data from SystemX) can be ported to another system under two assumptions: the new system uses industry-standard monitors such as Bro to output events that have the same semantics as in $\mathcal{E}$ and
a similar number of users. The assumptions reflect a limitation of the current PULSAR implementation, not its inference algorithm. While a new environment might have new events that are not in our defined event set $\mathcal{E}$, incorporating such new events into PULSAR would not require a retraining of the old FFs. Instead, only new FFs that are related to the new events need to be learned and used for inference.

\vspace{-3px}
\section{Related Work} \label{sec:related}
The signature-based techniques (SBTs)~\cite{paxson1999bro, bray2008ossec, egele2006using,shimamura2006using,massicotte2011analysis,zhaoowl} identify
specific hashes of attack payloads. Hence, they are suitable only for identifying a specific stage
of an attack (such as the use of a known rootkit at a late attack stage). They are not suitable for
preemptive detection of APTs because each stage may look benign when analyzed in isolation, but
together they show malicious intent (e.g., see \cref{sec:attackers_perspective}). The anomaly-based techniques (ABTs)~\cite{huang2013network,anceaume2014anomaly,xu2016sharper,yen2013beehive,du2017deeplog,shu2015formal,athreyaposter,le2012doubleguard,chu2012alert,dhakal2017machine,zoppi2016context,vadursi2016system,cinque2017entropy,feng2017multi,lee2017athena,zhang2011detecting,narayanan2018learning} construct a normal usage profile from past training data and measure a statistical
distance to find anomalous usage profiles at runtime. As a result, potentially novel attacks can
be captured, but at the cost of a high number of false detection~\cite{gates2006challenging}. Thus, security operators must select anomaly features carefully.

Detection models based on PGMs~\cite{holgado2017real,ning2003learning, teddi, cao2015preemptive}
have been built based on expert-defined libraries of known attacks. Thus, the detection accuracy is often
proportional to the manual work involved in creating features. \emph{Alert correlation}~\cite{broosquery,valeur2004comprehensive} and \emph{data provenance}~\cite{ji2017rain,hassan2018towards} techniques combine related events of the same attack instance into a stream of events. However, for large-scale systems, hundreds of events occur concurrently, so simply filtering and ordering events do not work.
\emph{Attack graph} techniques \cite{albanese2012time,sheyner2003tools,ravindranath2009change,nostro2014security,han2017work, didona2015using} illustrate possible attack scenarios, so they are useful for system administrators in deploying appropriate security monitors to defend their systems. The common problem across ML techniques~\cite{xie2010using,McHugh:2000:TID:382912.382923,du2014probabilistic,ravindranath2009change,lee2005scalable}, however, is that they have been trained on outdated~\cite{darpa} or synthetic datasets that contain artificial background traffic. Some attack detection techniques in~\cite{zhao2018owl,xu2016sharper} only consider host-level events. Forensic analyses~\cite{houmansadr2011cloud,pecchia2011identifying, pei2016hercule,yen2014epidemiological,vadrevu2017enabling,perdisci2006alarm,hu2016baywatch,yu2012filtering,akrout2014automated} localize compromised hosts post-incident, e.g., using post-compromise communication traces~\cite{oprea2015detection}, therefore, they are not suitable for preemptive detection. Overall, it is unclear whether these solutions can be successful in preempting APTs.

Applying ML and security presents several challenges~\cite{sommer2010outside}. First, while some ML models offer low FPR (e.g., 0.1\%), these models suffer from high FPR for large-scale deployment (e.g., a million security events per day $\times$ 0.1\% $=$ 1,000 false alerts per day). PULSAR lowers the FPR by only using statistically significance FFs learned from our real-world study of 120 APTs, thus compensating for the shortcomings of IDSs~\cite{demir2017towards}. In the future, it will likely be beneficial to deploy PULSAR on smart IDSs~\cite{zonouz2010cost}. Second, while existing work uses outdated~\cite{darpa,ozgur2016review}, short-term~\cite{DBLP:journals/corr/EsheteGHMSSVW16}, or limited variety of attack type datasets~\cite{algaith2017diversity,gouveia2017systematic}, our source training data are longitudinal (representing 10-year) and have been collected in real production traffic with up-to-date attack activities (until 2018). Third, while deep-learning models, e.g.,~\cite{du2017deeplog}, are promising, they offer analysts few explanations on how the models work. In contrast, PULSAR clearly shows the evolution of each attack stage; thereby enabling APTs response.

\section{Conclusion}\label{conclusion}
This paper presented PULSAR, a preemptive intrusion detection and response framework for the detection of Advanced Persistent Threats. PULSAR has been deployed at SystemX and demonstrated accurate attack preemption with a low FPR.

\newpage
\pagebreak
{
    \balance
    \bibliographystyle{IEEEtran}
    \bibliography{IEEEabrv,references}
}

\end{document}